\documentclass{aa}

\usepackage{epsfig}
\usepackage{amssymb}

\def\etal{{\it et al.~}} 
\def\ie{{\it i.e.,~}}
\def\eg{{\it e.g.~}}  

\begin{document} 

        \title{A Global Jet/Circulation Model for Young Stars}
        \author{T. Lery \inst{1,2}
        \and
        R.N. Henriksen \inst{3}
        \and
        J.D. Fiege \inst{4}
        \and
        T.P. Ray \inst{1}
        \and
        A. Frank \inst{5}
        \and
        F. Bacciotti \inst{6}}

        \institute {
        Dublin Institute for Advanced Studies,
        5 Merrion Square Dublin 2, Ireland
        \and
        GRAAL, CC 72, Universit/'e de Montpellier II,
        F-34095 Montpellier Cedex 05, France 
        \and 
        Department of Physics,
        Queen's University, Kingston,
        ON K7L 3N6, Canada
        \and
        McLennan Labs,  Univ. of Toronto,
        60 St. George Street,  Toronto,  ON  M5S 3H8, Canada
        \and
        Department of Physics and Astronomy, 
        Univ. of Rochester, Rochester, NY 14627-0171
        \and
        Osservatorio Astrofisico di Arcetri,
        L.go E.Fermi 5, 50125 Firenze, Italy
}

\date{ accepted 04/03/02 }

\titlerunning{A Global Model for YSO flows}
\authorrunning{Lery \etal}

\abstract{
Powerful, highly collimated jets, surrounded by bipolar molecular  
outflows, are commonly observed near Young Stellar Objects (YSOs).  
In the usual theoretical picture of star formation,  a jet is 
ejected from a magnetized accretion disk, with a molecular outflow 
being driven either by the jet or by a wider wind coming from the disk.  
Here, we propose an alternative global model 
for the flows surrounding YSOs. 
In addition to a central accretion-ejection engine driving the jet, 
the molecular outflow is powered by the infalling matter and follows 
a circulation pattern around the central object without necessarily 
being entrained by a jet.   
It is shown that the model produces a heated pressure-driven 
outflow with magneto-centrifugal acceleration and collimation. 
We report solutions for the three different
parts of this self-similar model, i.e. the jet, the infalling 
envelope and the circulating matter that eventually forms the 
molecular outflow.
This new picture of the accretion/outflow phase provides a
possible explanation for several observed properties of YSO 
outflows. 
The most relevant ones are the presence of high 
mass molecular outflows around massive protostars, and a realistic  
fraction (typically 0.1) of the accretion flow that 
goes into the jet.
   \keywords{Stars: formation --
             Methods: analytical --
             Magnetohydrodynamics (MHD) -- 
             ISM: jets and outflows  
               }
}

\maketitle


\section{Introduction}

Star formation occurs in molecular clouds. Observations have 
shown that the accretion phase, during which the central object 
builds up its mass, is very often accompanied by the powerful 
ejection of prominent bipolar outflows, that are comprised of fast 
jets (\cite{Ray96,eis00,reip01}) and slower and less collimated molecular 
outflows (\cite{Bachiller96}). A topic of great theoretical 
and observational interest is the dynamical connection between 
the jets and the molecular outflows. 
Strong arguments, such as the large masses observed 
in the CO flows, suggest that most of the molecular outflow material 
consists of swept up ambient gas (\cite{Bachiller99}).
The only possible source for this gas is the surrounding ambient cloud.
This implies that a driving agent accelerates molecular cloud material 
and gives rise to the observed CO lobes. 
An obvious candidate for such  an agent is  
the central fast jet.
The dynamical relationship between the  optical jet
and the surrounding bipolar molecular
outflows, however, is controversial
(\cite{snell,rodrig,bally,welch}).
At the heart of the controversy are the following 
questions. Has the central jet or wind sufficient thrust 
to drive the molecular outflow? If yes, how 
is the momentum transferred to the CO lobes?

There seems to be a consensus that jets have a magneto-centrifugal 
origin, and they are either launched from the accretion disk  
({\em disk wind}) (\cite{bp82,PP,WK,ST,PC,Ferreira,kp00,CS})
or from the location of the interaction of the protostar's magnetosphere  
with the disk ({\em X-wind}) (\cite{Shu94,Ferr00,shu00}). 
In the {\em disk wind} model, the engine consists of a 
Keplerian disk threaded by a magnetic field that is either
generated {\em in situ}, or advected in from larger scales, 
while {\em X-winds} are magnetised stellar winds 
where the interaction of a protostar's magnetosphere with the 
surrounding disk results in the opening of some magneto-spheric 
field lines. On the other hand,
the precise mechanism generating molecular outflows is poorly understood.
The latter are generally believed to be driven by a jet 
(\eg Padman \etal 1997, \cite{Chernin91})
or by a wide-angle wind (\cite{Barral,Smith,Shu91}), but all the
proposed possibilities present difficulties (\cite{Cabrit92}).  
For example, in the case of a wide-angle wind with a steady wind-blown 
cavity, the shocked wind converging at the tip of the cavity should
form a jet which is not observed. However, in the case of the wide-angle 
{\em X-wind} model, Shang, Shu \& Glassgold (1998), for example,
 have shown that a Herbig-Haro jet appears along the rotation axis,
\ie, in the central part of the cavity. On the other hand, for such 
models, the ambient medium has a density that decreases like 
$r^{-2}$ away from the core, which cancels the decrease in wind 
ram pressure due to the radial expansion, 
ensuring a constant shell velocity. This, however, does not seem
to be consistent with the shape observed for the 
largest flows (\cite{Cabrit92}).

In the case of jet-driven molecular outflow models, or 
``prompt entrainment'' mechanism,
the bulk of the molecular outflow is accelerated ambient gas near the
head of the jet or more precisely along the wings of its associated 
bow shock (\eg Chernin \& Masson 1995). 
This idea is supported by some recent simulations
(Micono \etal 2000), where it is shown that
a large fraction of the initial jet momentum can be transferred 
to the ambient medium. However, in some cases,  the jet may fail 
to provide the necessary momentum, as in L1551 IRS 5 (\cite{Fridlund}). 
Moreover, observations frequently 
suggest that jets decelerate near their outer edges, probably
due to entrainment of the surrounding material, but show 
little or no deceleration near the flow axis (\cite{Giovanardi}).
Also, in some cases, the axis of the cavity differs
from the axis of the jet, \eg by about 20$^{\circ}$ 
for HL Tau (\cite{Close489}),
or by about 40$^{\circ}$ in the case of NGC 2261 (\cite{Close489,Canto}).
Another major problem appears to be that steady jets produce 
molecular lobes with large aspect ratios (\cite{Chernin91}).
This would mean that the jet should be changing its position 
over time to accelerate different parcels of molecular gas 
(``wandering jet''), or varying its flow velocity 
resulting in internal shocks which push material sideways.
Finally, based on numerical simulations, the efficiency of 
prompt entrainment in accelerating molecules is found to be 
poor in the case of YSO jets with densities comparable to 
their ambient environment (\cite{downes}).

In addition to these problems, many bipolar outflows from 
massive protostars transport masses largely exceeding those of their 
associated stars. The bipolar outflow masses appear to range 
from about 10 to 4800 solar masses, with a mean value 
around 130 (Churchwell 1997). This probably indicates that 
much of the outflowing molecular gas could not have been 
accelerated close to the central YSO, since the protostar 
only represents a small fraction of the outflowing mass. 
Most of the models, discussed previously, face difficulties 
in explaining such massive outflows. Indeed, in this case, 
it seems likely that {\em deflection} 
of infalling matter into bipolar outflows may be a crucial mechanism, 
since entrainment and swept-up mass do not appear to be able to account 
for the very large observed outflow masses.
This is the mechanism that we propose in the present article.

Finally, 
we point out that a correlation exists 
between the bolometric luminosity of the 
central exciting source and the rate of momentum injection 
into the outflow (\cite{bally,Cabrit}).
These quantities are roughly consistent with a single power law
(with slope of about 0.7-0.8) 
across the full range of source luminosities extending over 
five orders of magnitude (\cite{Cabrit}). This continuous trend
suggests that similar mechanisms may be responsible for the 
production of molecular outflows from both low and high mass stars. 
As we will show, the present model can accommodate
this observational constraint.

We start by presenting the global model
in section~\ref{sec:mod},
and by writing down our set of self-similar MHD equations
in section~\ref{sec:equa}.
Then, in section~\ref{sec:sol},
we show the properties of the solutions along the streamlines, 
while, in section~\ref{sec:phys}, we discuss 
the main quantities that characterize the flow such as the 
fluxes of mass, momentum and energy, the plasma beta or the 
electric current. 
In section~\ref{sec:discussion} we 
present some physical consequences of the model
and we discuss several observational 
implications. Finally, we give our conclusions in 
section \ref{sec:concl}.  

\section{The Model \label{sec:mod}} 

\begin{figure}
\psfig{figure=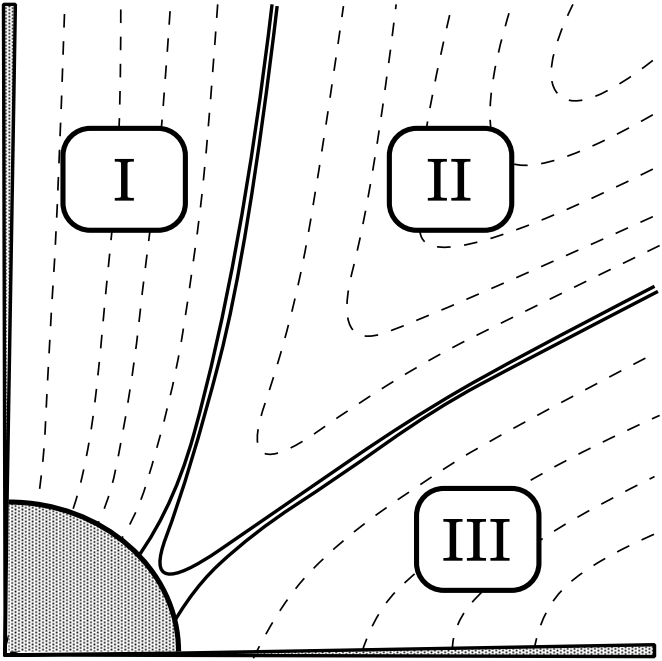,width=\linewidth}
\caption[ ]{
{\bf The different parts of the model}. 
Schematic representation of streamlines of the flows around YSOs
in the poloidal plane. The zones respectively correspond to 
(I) the jet, (II) the circulation region, and (III) the infalling 
envelope.  The thin accretion disk, the axis as well as 
the innermost region ($r<r_o \approx 100 AU$) of the protostar 
containing the accretion-ejection engine are represented in 
grey and are not treated in the present model.}
\label{fig:cart0}
\end{figure}

We propose a global model in
which the central jet is driven by a 
magneto-centrifugal accretion-ejection engine, 
while the molecular outflow is powered by the infalling 
matter through a quadrupolar circulation pattern around the central 
object. The situation is
schematically represented in Fig.~\ref{fig:cart0}. 
The molecular outflow may still be affected by 
entrainment from the fast jet, but this would be limited to the
polar regions and it would not be the dominant factor for its acceleration. 

The global model combines a jet model (I in Fig.~\ref{fig:cart0}), 
a circulation model (II) (\cite{HVG,FH,LHF,AFHL})
and an infalling envelope model (III). 
We suggest that the gas in all the three regions I, II, and III
can be described by the self-similar  
heated, quadrupolar and axisymmetric magnetohydrodynamic  
model (\cite{HVG,FH,LHF}), described here in section~\ref{sec:equa}.
In fact, it is possible to incorporate the jet model {\em inside} the  
self-similar circulation model, and to include the infalling envelope 
model near the equator. 
The torus and the jet were {\it not} part of the 
previously published models of this type (\cite{FH,LHF}), 
and their inclusion is a new result reported here.
Therefore {\em a description of the flows 
around YSOs can be obtained  from the axis to the equator with a 
single set of equations 
written in the framework of a self-similar hypothesis}.

Note that the model does not directly treat the 
thin accretion disk near the equator 
nor the accretion-ejection engine that provides the jet
in proximity of the central source, since the physical processes 
governing these regions are substantially different from the 
large-scale infall/outflow region.

\subsection{The Circulation Model} 

The circulation model (II in Fig.~\ref{fig:cart0}) 
has already been studied in
\cite{LHF}, and we only recall its main features here.
In the infall phase, the flow is slowed down by the increasing radial  
pressure gradients due to heating by the central source, and by an  
increase with proximity  to the central object 
of the strength of the  
barrier due to the rotation and the magnetic field. 
This pressure `barrier' deflects and accelerates, by means of its poloidal 
gradient, much of the infalling matter  into an axial outflow.
The Poynting flux included in the model increases both the velocity 
and collimation of the outflows by helping to transport mass and 
energy from the equatorial to the axial regions (see \cite{LHF} for details). 
The outflow velocity is 
naturally more pronounced as the axis of rotation is approached 
since the pressure gradients are strongest there and 
the density is smallest. These gradients 
also act to collimate the flow, forming a ``throat'' in the axial region. 

The main attraction for the circulation 
model is that large quantities of neutral material can 
be accelerated directly into the outflow without the need 
for jet entrainment, whose plausibility has long been debated.
Moreover, the model can generally explain 
the observed opening angles and velocity 
structure seen in high mass systems (\cite{FH,LHF}). 

Self-similar models cannot be globally consistent in 
time even if they are non-stationary, since they are, in general, 
ignorant of the initial conditions.
The self-similar circulation model  
is intermediate in space 
between a region around the symmetry axis containing the jet,
and the infalling envelope close to the disk.
These regions are described in the following. 
Note that the axis itself and the equatorial plane are strictly excluded 
from the domain of self-similarity although they
may be approached asymptotically.

\subsection{The Jet Model} 

In most jet models, the matter is thought to be transferred
from an accretion disk and/or a thick torus surrounding 
the central forming star, to the jets.
In the most widely accepted scenario  
the outflow acceleration is mediated by magnetic fields that thread
Keplerian type disks. In the present model, no detail is given about the
origin of the jet itself but our description is compatible with most
models mentioned in the introduction. We consider that the system consisting
of the protostar and the thin accretion disk acts as a reservoir 
for the jet and only study the flow inside the jet 
(I in Fig.~\ref{fig:cart0}). The same 
set of equations used for region II can be applied 
to the jet. As for the circulation solutions, 
the velocity is larger closer to the axis but
the dynamics is dominated by the variations
of the various quantities in the radial direction, \ie along the jet. 
The jet model applies only to a limited range of finite radii, where
the self-similarity applies.
Specifically, our model cannot address the launching mechanism of the
jet near the star.  The structure of a supersonic jet far from 
the source is disconnected from its origin.  Thus, it is not essential 
to include a detailed description of the central engine in our model.

\subsection{The Self-similar Infalling Envelope Model}

During the formation of the star, it is expected that an accreting 
disk forms and feeds the central protostar from infalling ambient 
cloud core material; these disks may have radii from 20 AU 
up to several hundred AU, and masses containing a 
fraction of the central protostar mass.
Hence, the last element of the model is 
the region where material is falling from the cloud 
onto the central object.
It consists of a standard flattened dense accretion 
disk, not treated by the present model 
(\ie the equatorial-most region in Fig.~\ref{fig:cart0}), 
and of an extended and less dense torus of infalling gas and dust
(III in Fig.~\ref{fig:cart0}).
The latter can be as large as a few thousand AU and is also described 
by the set of self-similar equations of the circulation model.
The material, for both the circulation and the infalling envelope
models, originate from the same region in the molecular cloud.
We thus suggest that the same set of equations may apply in both cases.
This statement would not be applicable to the thin disk 
where resistivity, ionization fraction and other parameters 
can substantially differ from the outer regions.

\section{The Equations of the Model \label{sec:equa}} 

\subsection{The Variables}

The solutions are developed within the context of 
$r$-self-similarity wherein a power of $r$ multiplies 
an unknown function of $\theta$; the spherical coordinates 
$r$, $\theta$ and $\phi$ being used.
The only physical scales that enter into our calculation
are the gravitational constant $G$, the fixed central mass $M$, 
and a fiducial radius $r_o$ (\cite{FH,LHF}).
Note that Fiege \etal expressed $r_{o}$ in terms of $G$, $M$,
and the central luminosity $L_\star$, while Lery \etal
offered a different choice. In the present work,
we leave $r_o$ as a scale that can be freely specified for
any observed outflow source.
We also introduce two other parameters that are the limiting 
angles between the jet, the infall and the circulation zones.
The first angle, $\theta_{min}$, goes from the axis of
rotation to the boundary of the jet (angle between zone I and II 
in Fig.~\ref{fig:cart0}). The second angle,
$\theta_{max}$, corresponds to the limit of the circulation
region (angle between zone II and III in Fig.~\ref{fig:cart0}).
The power laws of the self-similar system are determined, up to a single 
parameter $\alpha$, if we assume that the local gravitational field is 
dominated by a fixed central mass. 
In terms of the fiducial radial distance, $r_o$, the 
self-similar symmetry is sought as a function of two scale 
invariants, $r/r_o$ and $\theta$, in a separated power-law 
form. The self-similar index $\alpha$ is a free  parameter 
of the solution, but must lie in the range $-1/2 < \alpha \le 1/4$,
for simultaneous infall and outflow to occur.
The radiation field can help us to understand the 
physical meaning of $\alpha$. Indeed,
in this context, the self-similar behaviour 
that we impose on the radiation field takes the form
$\vec F_{rad}=\left({GM}/{r_o}\right)^{3/2}
{M}/{r_o^3}\left({r}/{r_o}\right)^{\alpha_f-2}\vec f(\theta)$
(\cite{FH,LHF}).
In this equation, the index $\alpha_f$ is a measure of 
the radiation energy balance as a function of radial distance. 
When $\alpha_f$ is negative, it corresponds to a loss in 
energy due to the net radiation leaving the region, 
while, for a positive $\alpha_f$, the system has an energy 
input due to the absorption of radiation. If one supposes that 
the opacity is predominantly due to dust,  the index 
$\alpha_f$ is related to $\alpha$  
by $\alpha_f\approx -2 (1/4-\alpha)$.
Consequently, the case $\alpha_f=0$, where there is no net flux
of radiation through the domain of our solution, 
corresponds to $\alpha=-1/4$. For larger (respectively 
smaller) values of $\alpha$, there is a net loss (gain) 
of energy by radiation (\cite{FH}).

Hence, if we assume that the gravitational 
potential  is dominated by the central mass,
\ie self-gravitation is negligible, the equations of radiative MHD 
admit the following radial scaling relations 
for the variables
\begin{equation}
{\bf v} = \sqrt{\frac{GM}{r_o}} \left(\frac{r}{r_o}\right)^{-1/2} 
\ {\bf u}(\theta),
\label{self1}
\end{equation}
\begin{equation}
\rho = \frac{M}{r_o^3} \left(\frac{r}{r_o}\right)
^{2\alpha-1/2} \mu(\theta),
\end{equation}
\begin{equation}
B_{\phi,p} = \sqrt{\frac{G M^2}{r_o^4}} \left(\frac{r}{r_o}\right)
^{\alpha-3/4} \frac{u_{\phi,p}(\theta)}{y_{\phi,p}(\theta)},
\end{equation}
\begin{equation}
p = \frac{G M^2}{r_o^4} \left(\frac{r}{r_o}\right)
^{2\alpha-3/2} P(\theta),
\end{equation}
\begin{equation}
\frac{k T}{m_\mu m_H} = 
\frac{G M}{r_o} \left(\frac{r}{r_o}\right)^{-1}
\ \Theta(\theta),
\label{self2}
\end{equation}
\begin{equation}
\vec F_{rad}=\left(\frac{GM}{r_o}\right)^{3/2}
\frac{M}{r_o^3}\left(\frac{r}{r_o}\right)^{\alpha_f-2} \vec f(\theta) ,
\label{selfrad}
\end{equation}
where ${\bf v}$, $\rho$, ${\bf B}$, $p$, $T$ and 
${\bf F}_{rad}$ respectively correspond
to the velocity, the density, the magnetic field, the pressure, the
temperature and the radiative flux.
In these  equations the microscopic constants are represented by
$k$ for Boltzmann's constant, $m_\mu$ for the mean atomic weight,
$m_H$ for the mass of the hydrogen atom. 
In the last equation, the index
$\alpha_f$ is a measure of the loss (if negative) 
or gain (if positive) in energy by radiation
as a function of radial distance. 
The self-similar variable directly related to 
magnetic field $y$ can be divided into poloidal
and toroidal components. In the present model the two components
are not equal. They are respectively defined by
$y_{p,\phi}(\theta)=M_{ap,a\phi}/\sqrt{4\pi\mu(\theta)}$.
Consequently the system also deals with two different components of the 
Alfv\'enic Mach number $M_{ap}$ and $M_{a\phi}$ defined by
$M^2_{ap,a\phi}(\theta)\equiv
{\vec{v}_{p,\phi}^2}/\left({\vec{B}_{p,\phi}^2/4\pi\rho}\right)$.
We will refer to $\Theta_0$ as the value of the self-similar 
temperature $\Theta(\theta)$ on the axis.

\subsection{The Equations}

In order to make the system tractable, we 
assume axisymmetric flow so that $\partial/\partial\phi=0$ and all flow 
variables are functions only of $r$ and $\theta$. 
We further restrict ourselves to steady models 
(\ie $\partial/\partial t=0$).
Magnetic field and streamlines are required to be quadrupolar  in the 
poloidal plane for the circulation model
(see \cite{LHF} for the specification of the 
boundary conditions). Under these assumptions 
the self-similar equations are:
\begin{enumerate}
\item {\underline {Mass flux conservation}}
\begin{equation}
\left (1+2\,\alpha\right )\mu\,u_{{r}}+
\frac {1}{\sin{\theta}}
\frac{d}{d\theta}\left (\mu\,u_{\theta}\sin{\theta}\right )
=0 ,
\label{MFC}
\end{equation}
\item{\underline {Magnetic flux conservation}}
\begin{equation}
{\frac {\left (\alpha+5/4\right )u_{{r}}}{y_{{p}}}}+
\frac {1}{\sin{\theta}}
{\frac {d }
{d \theta}}\left ({\frac {u_{\theta}\sin{\theta}}{y_{{p}}}}
\right )
=0 ,
\label{BFC}
\end{equation}
\item{\underline {Radial component of momentum equation}}
\begin{eqnarray}
u_{\theta}\frac {du_{r}}{d\theta}
\left (1-{M^{-2}_{{ap}}}\right )-\left ({u^2_{\theta}}+{u^2_
{\phi}}\right )\left (1-{\frac {\alpha+1/4}{M^2_{ap}}}
\right )
\nonumber 
\\
-\frac{u^2_{r}}{2}
-\left (\frac{3}{2}-2\alpha\right )\Theta+1+
\frac {u_{r}u_{\theta}}{M^2_{ap}y_{p}}\frac {dy_{p}}{d\theta}=0,
\end{eqnarray}
\item{\underline {$\theta$-component of momentum equation}}
\begin{eqnarray}
\frac{u_{r}u_{\theta}}{2}
\left (1-{\frac {2\,\alpha+1/2}{{M^2_{ap}}
}}\right )
+{u^2_{\phi}} \cot~\theta
\left ({M^{-2}_{a\phi}}-1\right )
&
\nonumber 
\\
+{\frac {u_{r}}{{M^2_{ap}}}}{\frac {du_{r}}{d\theta}}
+u_{\theta}{\frac {du_{\theta}}{d\theta}}
+{\frac {u_\phi}{M^2_{ap}}}{\frac {du_\phi}{d\theta}}
-{\frac 
{{u^2_r}}{y_p M^2_{ap}}}{\frac {dy_{p}}{d\theta}}
&
\nonumber 
\\
-{\frac {
{u^2_\phi}}{{y_\phi M^2_{a\phi}}}}{\frac {dy_\phi}{d\theta}}
+{\frac {d\Theta}{d\theta}}
+{\frac {\Theta}{\mu}}{\frac {d\mu}{d\theta}}
=0 , &
\end{eqnarray}
\item{\underline {Angular momentum conservation}}
\begin{eqnarray}
\frac{1}{u_{\phi}}
{\frac {du_{\phi}}{d\theta}}
\left (1-\frac {1}{M_{ap}M_{a\phi}}\right )
+\frac{u_r}{u_{\theta}}
\left (1/2-{\frac{\alpha+1/4}{M_{ap}M_{a\phi}}}\right)
\nonumber 
\\
+\cot~\theta
\left (1-{\frac {1}{M_{{ap}}M_{{a\phi}}}}\right )+{
\frac {1}
{y_{{\phi}}M_{{ap}}M_{{a\phi}}}}
{\frac {dy_{\phi}}{d\theta}}
=0 ,
\end{eqnarray}
\item{\underline {Faraday's Law plus zero comoving electric field}}
\begin{equation}
{\frac {d (u_{{\phi}}u_{{\theta}})}{d \theta}}
+\left [\alpha-\frac{1}{4}\right ]u_{{\phi}}u_{{r}}+u_{{\phi}}u_{{
\theta}}{\frac {d }{d \theta}}\ln \left[\frac{1}{y_{{p}}}-\frac{1}{y_{{
\phi}}}\right]=0 .
\label{Faraday}
\end{equation}
\end{enumerate}
This set of equations can produce either circulating flows,
pure infall, or pure outflow. 
Note that the boundary conditions guarantee mass
conservation in the circulating region (II), while
in the two other cases, mass fluxes are not conserved.
Indeed, the source acts as a reservoir for the pure outflow (region I) 
while the accretion disk would act as sink for the pure infall 
(region III). Therefore, any discussion on net mass infall/outflow
can only apply to region I and III. 

\section{Example of Solution\label{sec:sol}}

\begin{figure}
\psfig{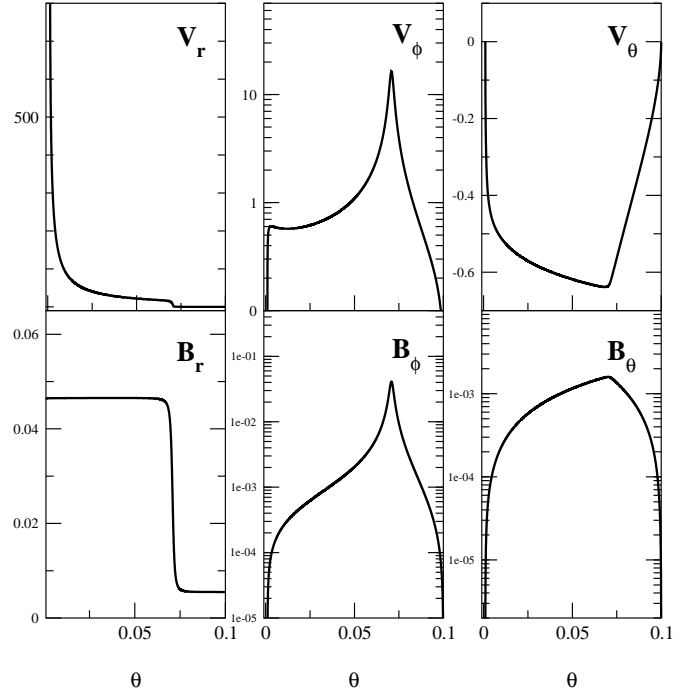}
\caption[ ]{
{\bf YSO Jet solution as a function of the angle from the axis}. 
Velocities and magnetic field are respectively
given in km s$^{-1}$ and Gauss.
Parameters are $\alpha=-0.12$, $\Theta_0=0.35$.
Here, in order to dimensionalize the quantities,
the mass of the central protostar is set to one solar mass,
and $r$ is kept fixed at a distance of $10^4$ AU
from the source.
}
\label{fig:soljet}
\end{figure}

From a numerical point of view, we integrate the set of equations 
from a given solution close to the axis up to a given angle $\theta_{min}$. 
This solution near $\theta=0$ has finite values for $V_r$, 
$B_r$ and the density. The other components of the velocity and 
magnetic field are vanishing, otherwise they would show jump discontinuities
at the boundaries. The last point from the jet model solution 
can be used as initial condition for the integration of the same set 
of equations of the circulation model. However, we have restricted 
ourselves to solutions 
in which the boundary conditions for the circulation region 
are zero for  $B_\phi$, $B_\theta$, $V_\phi$, $V_\theta$, a good 
behaviour for the circulation solution.  
We note that under these hypotheses no material crosses the boundaries 
defined by the cones delimiting the three regions, and that the 
solutions are in pressure balance.

Once the circulation solution is obtained, we integrate then 
from this solution down to a given angle ($\theta_{max}$) 
in the equatorial region, excluding a small zone corresponding 
to the thin accretion disk.
We also ensure that $B_\phi$, $B_\theta$, $V_\phi$, $V_\theta$
vanish at the interface between the circulation solution and 
the infalling envelope. The main difference
between the two solutions is the change of signs
of the $\theta$-components of the velocity and the magnetic field.
This corresponds to a solution that goes towards the axis
in the case of the circulation solution, and towards the equator
for the infalling envelope solution.

The three components of the velocity and of the magnetic field 
for a jet solution are shown in Fig.\ref{fig:soljet} 
as a function of the angle between the axis and a limiting angle
($\theta_{min}=0.1$ in this case), 
and at a distance of $10^4$ AU, the mass of the central protostar 
being set to one solar mass. Radial velocities
of several hundred of km s$^{-1}$ are obtained near the axis. 
There, the radial component of the magnetic 
field also dominates the other component. 
The rapid increase in velocity towards the axis is easily 
explained as follows.  From Bernoulli's law, a decrease in 
density along a streamline, and hence in pressure, 
corresponds to an increase in velocity.
This is what we observe towards the axis, where material encounters a low
pressure region.  Bernoulli's law strictly applies only in the 
absence of a Poynting flux, in which energy is directly transferred 
between the magnetic field and the kinetic energy.  
However, it is clear from our solutions that the Poynting flux 
is not dominant in this region.
Note that the flatness of $b_r$ in the innermost part of the jet
is not a general behavior of the self-similar solutions 
but corresponds to the present set of parameters.
It is noteworthy that the strength of the radial component 
of the field could help to stabilize the jet w.r.t. internal 
instabilities (\cite{lb}). In most of the jet, the angular 
momentum and the magnetic pressure are negligible
with respect to the kinetic energy, whereas, close to the 
interface with region II, the hoop stress becomes stronger,
due to the increase in $B_\phi$. The change of slope in the 
variables at $\theta \sim 0.07$ corresponds to the location 
of the steep gradient in $B_r$ and of the maximum 
amplitude of $B_{\phi}$ in the jet. If this angle is taken 
as a measure of the broadening of the jet during the  
propagation at large scales, we may derive a full opening 
angle of the flow of about 8$^{\circ}$, which is in the 
range of values derived from the observations (\cite{eis00}).

\begin{figure}
\psfig{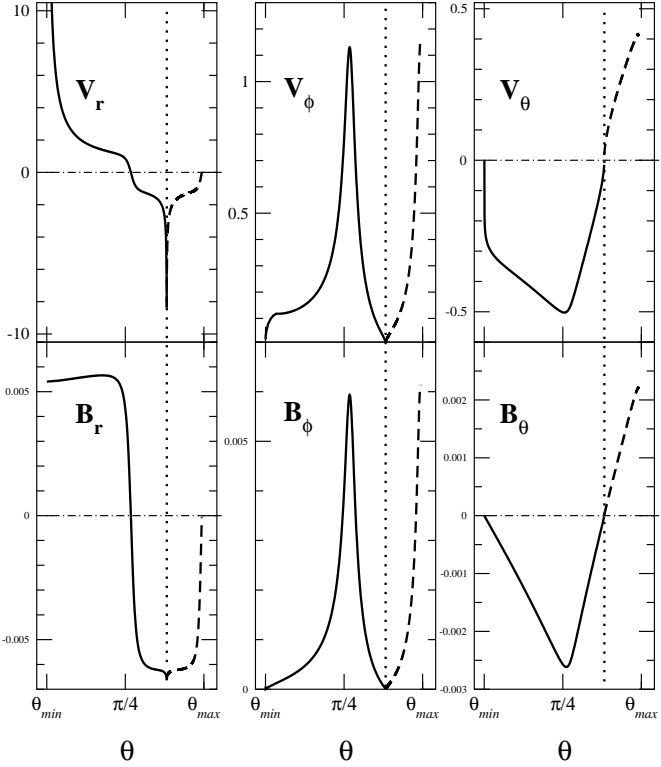}
\caption[ ]{
{\bf Circulation and infalling envelope solutions}. 
Solid lines and dashed lines respectively correspond to
the circulation and the infall solutions.
Velocities and magnetic field are respectively
given in km s$^{-1}$ and Gauss.
The distance from the source is taken to be $10^4$ AU.
Parameters for the solution are $\alpha=-0.12$, $\Theta_0=0.35$,
$\theta_{min}=0.1$ and $\theta_{max}=1.55$, the transition between 
the solutions being at $\theta=1.2$.
}
\label{fig:sol}
\end{figure}
The solutions in the circulation and the infalling envelope
torus region are reported in Fig.\ref{fig:sol}. 
Following the flow, the radial velocity is negative 
in the infalling region, $\theta_{tp} > \pi/4$,
it changes signs at the turning point (the closest point 
to the source) located at $\theta_{tp}\approx \pi/4$ 
in the present case.  Finally it increases and passes from
a few km s$^{-1}$ to more than 10 km s$^{-1}$ in the 
axial region. 
Note that the radial velocity 
vanishes in the equatorial region where it connects to the thin
accretion disk, for this solution. 
This is neither a general result, nor a boundary condition
that we impose. Solutions with negative
values of the radial velocity can also be obtained.
The rotation is maximum 
at the turning point but it is still dynamically  important 
closer to the disk. The $\theta$-component of the 
velocity shows that the gas is moving towards the axial region
in the circulation model while it is directed towards the equator
for the infalling envelope. 
Finally, the maximum of $V_\theta$ corresponds to
a maximum in the toroidal component of the magnetic field $B_\phi$ which 
produces a radial ``hoop-stress'' in the outflow that helps 
the collimation (\cite{leryfrank}). 

\begin{figure}
\psfig{figure=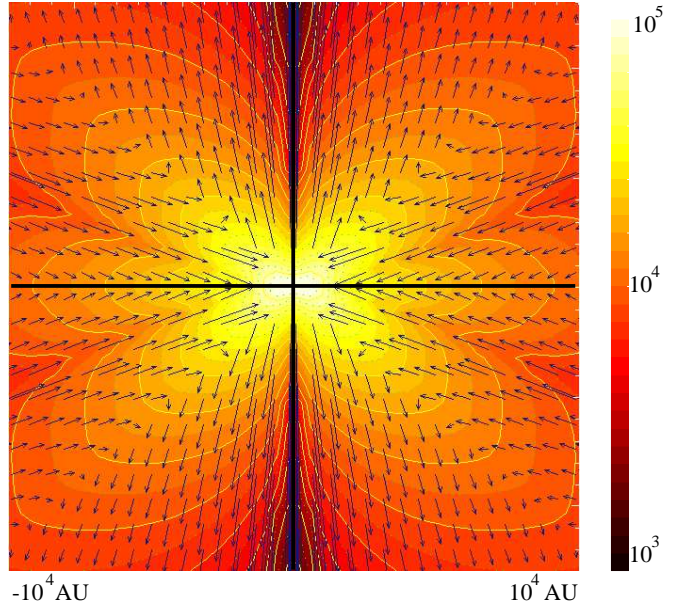,width=\linewidth}
\caption[ ]{
{\bf 2D iso-contours of the density and velocity field}. 
The protostar is located in the center. The density
ranges from $10^3$ to $10^5$ cm$^{-3}$,
the lowest densities corresponding to the darkest regions.
The size of the box is given in AU.
}
\label{fig:density}
\end{figure}
We also present the density iso-contours, 
with superimposed poloidal velocity vectors,  
in Fig.\ref{fig:density} for the entire model in the poloidal plane.
The density in the jet is lower than the density in the 
molecular outflow by at least a decade in the present 
example, but this is not a general result of the model.
In the molecular outflow the density increases 
from the outflow to the infalling region, decreases at the interface
and reaches again a maximum near the equator. 
Other examples of density profiles as well as 
variations of the main quantities along a streamline
can be found in Lery \etal (1999) for the circulation model.

\begin{figure}
\psfig{figure=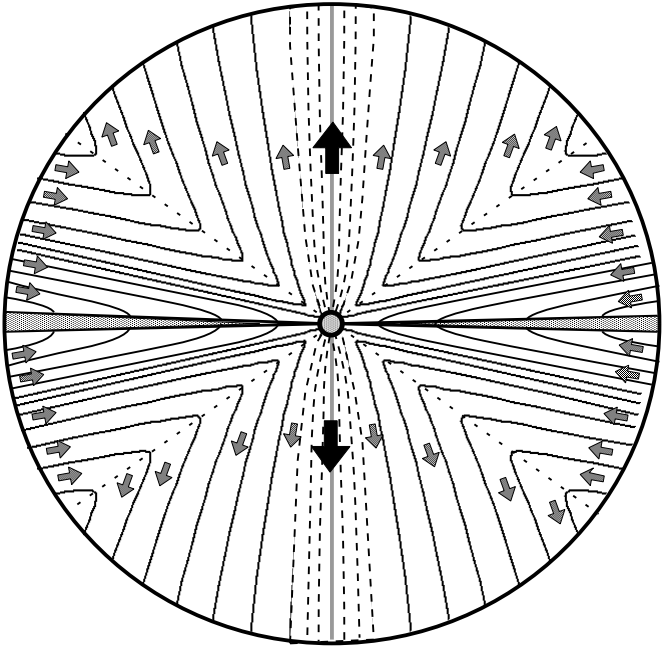,width=\linewidth}
\caption[ ]{
{\bf Streamlines around a one solar mass  protostar}. 
Streamlines projected in the poloidal plane and given by the solution
for $\alpha=-0.12$, $\Theta_0=0.35$. 
The solid lines correspond to the circulation 
and the infall models, and the dashed lines to the jet model.
The arrows indicate the direction of the gas on the streamlines. 
The typical sizes of the solution is $10^4$~AU.
}
\label{fig:streamline}
\end{figure}

We finally present the streamlines obtained with the solution
given above  in Fig.~\ref{fig:streamline}. The streamlines are 
projected in the poloidal plane and are shown with arrows that 
indicate the direction of the flow. 
The outermost region is dominated by the circulation 
solution except close to the axis and
in the equatorial region. On the other hand, the 
accretion-ejection is dominated by the dynamics near the protostar.

\section{General properties of the Global Model}\label{sec:phys}

\subsection{The Fluxes}\label{sec:fluxes}

The empirical relationships between the mass, momentum, and 
energy fluxes provide important observational constraints on
physical models of bipolar outflows.
Indeed, in our model, the total mass flux $\dot M$, 
momentum flux $\dot P$ and kinetic energy flux $\dot E$
are respectively given by
\begin{equation}
\dot M\propto\left({r}/{r_o}\right)^{1+2\alpha},
\dot P\propto\left({r}/{r_o}\right)^{1/2+2\alpha},
\dot E\propto\left({r}/{r_o}\right)^{2\alpha}.
\end{equation}
It follows that models with
$\alpha<-1/4$ and $\alpha>-1/4$, \ie with loss 
in energy by radiation through the boundaries, 
have mass and momentum fluxes dominated by 
the inner and outer regions respectively. 
Moreover, the energy flux is dominated
by the interior region if $\alpha<0$ (and by the exterior
for $\alpha>0$). 

Let us now consider in more detail the mass fluxes in this solution.
The total mass flux 
coming from the molecular cloud  is given by
$\dot M_{in}=\dot M_{infall}+\dot M_{acc}$,
where $\dot M_{infall}$ corresponds to the infalling portion 
of the circulation pattern, 
and  $\dot M_{acc}$ to the mass flux in the envelope which 
is subsequently channeled  toward the star
through the accretion disk.
The total mass loss rate reads
$\dot M_{out}=\dot M_{MO}+\dot M_{jet}$,
$\dot M_{MO}$ and $\dot M_{jet}$ being  the mass loss 
rate of the molecular outflow and of the jet, respectively
Since the infalling gas  is conserved along 
the path in the circulation model,
$\dot M_{infall}=\dot M_{MO}$. Thus,
the global mass flux conservation reads 
\begin{equation}
\dot M_{jet}=\dot M_{acc}-\dot M_{\star} = f~\dot M_{acc},
\label{mdot1}
\end{equation}
where $\dot M_{\star}$ is the accretion rate 
onto the star, and  $f$ is the fraction of the accretion flow that goes into
the jet.

The mass fluxes are calculated as
$\dot M=\int_{\theta_{in}}^{\theta_{out}} 
{ \rho(\theta) V_r(\theta) 2 \pi R^2 d\theta}$,
where $\theta_{in}$ and $\theta_{out}$ are the poloidal angles 
delimiting each region of the system,
and $R$ a fiducial distance in the radial direction. We thus consider 
fluxes through sections of a spherical surface centered on the star,
the distance of which from the source is arbitrary as long as the 
material keep flowing in conical volumes (far enough from the disk).
By using this method, the values that we obtain can be
tunable within a few orders of magnitude of
$G^{1/2}M^{3/2}r_o^{-3/2}$.
For the solution presented here, 
the mass fluxes in the jet, the molecular 
outflow and the accreting torus are found to be, respectively:
\begin{equation}
\dot M_{jet}\approx 6\times10^{-7} M_\odot yr^{-1},
\end{equation}
\begin{equation}
\dot M_{infall} = \dot M_{MO} \approx 8.5\times10^{-6} M_\odot yr^{-1},
\end{equation}
\begin{equation}
\dot M_{acc} \approx 6\times10^{-6} M_\odot yr^{-1}.
\end{equation}
The value of the mass loss rate in the jet is
of the same order of magnitude of the ones
 found in several observational studies, \eg  
$\dot{M}_{jet} \sim 3.7~10^{-7}$ for HH34 (\cite{Bacciotti99}),
$\dot{M}_{jet} \sim 2.4~10^{-7}$ for DG Tau (\cite{Bacciotti01}).
Our value for $\dot M_{MO}$ is also similar to the ones estimated
observationally for typical molecular flows (\cite{Bachiller96}).
We also get $f=\dot{M}_{jet} / \dot M_{acc} \approx 0.1$, for this
particular model, again in good agreement with observations 
and  predicted by other MHD models
specifically treating the launch of the jet. For example, 
$f\approx0.03$ for self-similar disk-wind models 
(\eg Ferreira 1997),
and $f\approx0.3$ for X-wind models 
(\eg Shu \etal 1994).

Our model only applies to the YSO environment from 
$r_o\approx 10^2-10^3$ AU to $10^4-10^5$ AU, which is 
outside the region where the jet is launched.
However, we speculate that
a quadrupolar jet could form on a smaller scale.
In order to understand the type of acceleration
in the jet, it can be useful to look at
$\nu=4 \pi \eta v_{\phi o} / B_o$, where
$\eta=\rho v_r/B_r$ is the mass to magnetic flux ratio.
In the example presented here, we have 
$\nu=0.03\ll 1$. This shows that the flow is
not accelerated magnetically but centrifugally or thermally.
This could be done 
through series of giant flares close to the
central object (\cite{feigelson}), 
but this is beyond the scope of the present work.

Finally we point out that in our model the poloidal 
momentum flux, calculated as
$\dot P=\int_{\theta_{in}}^{\theta_{out}} 
{ \rho(\theta) V^2_r(\theta) 2 \pi R^2 d\theta}$,
is about the same in the jet and in the molecular outflow. 
This is also in agreement with observations
(\cite{Bacciotti99}). In our model, however, there is no
strong dynamical relationship between the jet and the molecular outflow.
Thus it seems that the observational similarity between the momentum rates
in such outflows may have no relationship with the mechanism accelerating
the molecular lobes.  
In a future paper, we plan to constrain the global model 
with observations of specific sources for which the opening angle
as well as the mass fluxes can be estimated.

\subsection{The Energetics of the Flow along Streamlines}

The acceleration of a magneto-hydrodynamic wind is usually described 
in terms of the centrifugal force acting along the field lines. 
Alternatively, the development of the flow can be described in 
terms of energies. At the closest point to the source,
part of the energy is in the form of magnetic energy, which 
gets converted (in part) into kinetic energy, as we 
will see. In this section and in the next ones we will describe 
and study the variations of the energies, of the plasma beta and
of the electric current along streamlines in order to 
illustrate the physical aspects of the model. 
We will concentrate, in particular, 
on the circulation model that is the corner stone of the global 
model. 

\begin{figure}
\psfig{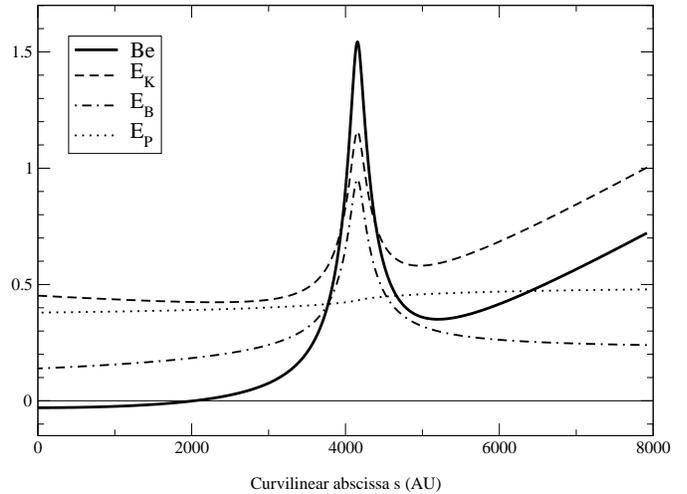}
\caption[ ]{
{\bf Energetics of the Flow along a Streamline}
The Bernoulli-like function $Be$ (heavy solid 
line) is represented with its various components, 
the specific kinetic energy $E_K$ (dashed line),
the specific magnetic energy $E_B$ (point-dashed line) 
and the specific enthalpy $E_P$ (dotted line)
in the circulation model. The peaks take place at the turning point.
The curvilinear abscissa $s$ along a streamline is given in AU
($\alpha=-0.12$,$\Theta_0=0.5$, $r_o \approx 100$AU, $M\approx 1M_\odot$)
\label{fig:Be5}
}
\end{figure}

In analogy to Bernoulli's theorem, it is useful for our 
discussion  to define a function $Be$
that corresponds to the sum of the specific kinetic energy, 
the specific potential energy (or gravitational energy), the 
specific enthalpy of the gas, and the specific magnetic energy, 
all divided by the local specific gravitational energy
to make $Be$ dimensionless. In the present case, the 
gravitational energy per unit mass is approximately
the square of the local free-fall velocity. One should note that 
$Be$ is not a streamline integral due to the presence of 
the magnetic field. Nevertheless, it remains useful for discussing
the flow of energy between the various components.
This function of $\theta$ is given by
\begin{eqnarray}
Be(\theta)&\equiv& (E_k +E_G + E_H + E_B)/ (GM/r) 
\nonumber 
\\
&=&(u_r^2+u_\theta^2+u_\phi^2)/2-1 +\Theta 
+ u^2/(8 \pi \mu y^2),
\end{eqnarray}  
where $\vec u$, $\Theta$ and $\mu$ respectively
represent the dimensionless velocity, temperature and density.
The variables $E_k$, $E_G$, $E_B$  respectively correspond
to the kinetic, gravitational, and  magnetic energies, while $E_H$ is 
the enthalpy. The variable $y$ is related to the magnetic field by 
$B \propto u/y$. The parameter $\alpha$
is the self-similar exponent, $M$ is the mass of the
central object and $r_o$ is a fiducial radius
(See paper I for details). 
Note that the full energy equation that has $Be$ as a first 
integral has been given previously (See equation 15 in paper I or
 Henriksen 1997).
In the flow, the positivity of $Be$
suggests that the gas should reach infinity with a net positive 
energy, although this is not strict since the energy 
is not conserved. On the other hand, when $Be$ is negative, 
the gas should not spontaneously escape to infinity. 
Note that the positivity of $Be$ does not imply a lack of 
conservation of energy, but the excess $Be$
represents an energy transfer from hotter regions at smaller 
radii to motion, \ie mainly from the enthalpy to the kinetic 
energy. Of course when proper boundary conditions are applied 
and the full global problem is solved, the total energy would 
be conserved. The net gain in $Be$ from 
the infall region to the outflow region in its 
various components (See Fig.~\ref{fig:Be5}) 
represents the energy transferred to the gas
when it is closest to the star.

In Fig.~\ref{fig:Be5}, the solutions are for
$\alpha=-0.12$,$\Theta_0=0.5$, $r_o \approx 100$AU, 
$M\approx 1M_\odot$. Note that $Be$ is smaller
than the kinetic energy in this figure because the gravitational 
energy, not represented here, is negative.
Far from the central object, the dominating
energy is gravitational at the beginning of the infall. 
There, repulsing forces such as the pressure barrier, 
the centrifugal force or the magnetic force are weaker 
than gravity. Therefore, $Be$ starts with a negative value.
When the gas gets closer, both the kinetic and magnetic 
energies increase. Then $Be$ becomes positive, which allows 
the gas to escape in the form of an outflow.
It seems most natural that  more powerful outflows preferentially 
form along the rotational axis since there $Be$ is most 
positive in all cases. Thus a bipolar morphology for outflows
arises quite naturally in the present scenario.
It is worth comparing this result with the zero-pressure limit 
(\cite{LHF}) where $Be$ is conserved on streamlines. In this case, 
the model does not produce velocities greater than the escape 
velocity anywhere. This is  due to the fact that energy that is 
first gained by the magnetic field at the expense of gravity and 
rotation during infall is subsequently returned as gravitational 
potential energy and kinetic energy in the outflow phase.

The most rapidly outflowing gas is always near the symmetry axis
because these streamlines pass closest to the star, 
deeper into the gravitational potential well. Also, the 
material on these streamlines is heated the most vigorously. We 
note that we do not include radiation pressure in our model.
In the present context, the role of the radiation produced by
the source is to transport outward thermal energy. 
The radiation is eventually absorbed contributing 
to the gas heating.

As the gas gets closer to the source, it rotates faster. 
The infalling plasma therefore has a larger electric 
current driven by the rotational motion. This increases 
the magnetic energy, which is eventually converted into 
kinetic energy as the gas is redirected outwards. 
The magnetic field acts to collimate and accelerate the gas 
towards the polar regions. There the flow presents 
a strong poloidal velocity and a very low 
magnetic energy. 

\begin{figure}
\psfig{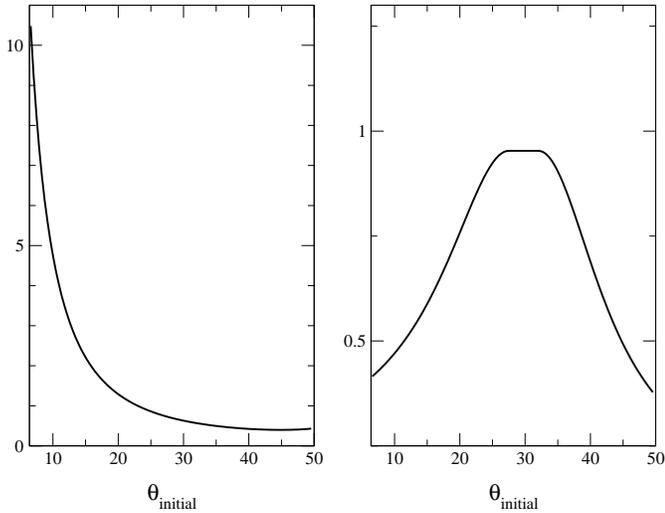}
\caption[ ]{
{\bf Gain in kinetic energy}
given by the ratio of the final to initial values in the flow,
as a function of the initial angle of infall (left panel).
Magnitudes of the kinetic energy
at the turning point as a function of the initial angle of infall 
(right panel)
(Same set of parameters as in Fig.~\ref{fig:Be5})
\label{fig:BeKP}
}
\end{figure}

In Fig.~\ref{fig:BeKP}, we show the 
ratio of the final to initial magnitudes of the kinetic energy
in the flow as a function of the initial angle 
of infall. The solution has been obtained for the same set of 
parameters as in Fig.~\ref{fig:Be5}.
Since the gas falls from the equatorial regions and 
ends up in the axial zone, the ratio gives the gain in energy 
acquired during the circulation. The gain in kinetic 
energy varies drastically. It goes from less than unity 
in the region far above the disk 
($\theta_{in}\approx 45^{\circ}$) to more than 10 for 
streamlines that start close to the equatorial 
region ($\theta_{in}\approx 10^{\circ}$). 
In the latter case, if the gas starts with an infalling velocity
around 5~km s$^{-1}$, it will acquire an escaping velocity of at least
50~km s$^{-1}$ (\cite{Bachiller99}). 
Again, the circulation is thus very 
efficient in accelerating the flow for the streamlines that start
the closest to the  equator, and then travel the closest 
to the central protostar, going deeper in the potential well. 

Therefore, in the axial region,
the velocity will appear to decrease as a function of the distance
from the axis. Such properties has been observed for collimated
molecular outflows (\eg Guilloteau \etal 1997), and more recently,
also for the central optical jet (\cite{Bacciotti00}).
On the right panel of Fig.~\ref{fig:BeKP}, the kinetic energy
is represented at the turning point
as a function of the initial angle of infall.
The energy is maximum for an intermediate
range of angles, between $25^{\circ}$ and  $35^{\circ}$.

Summarizing, the material ejected in the outflow 
acquires a positive energy due to a combination of
heating by the central protostar, pressure gradients developed 
by the infalling material, rotation and magnetic forces due to the
non-zero Poynting flux. The study of the energetics
along a streamline helps to describe how the local infalling 
flow is channeled into a bipolar outflow.

\subsection{The Electric Current and the Plasma $\beta$ Parameter}

The electric current is a major indicator of the collimation
of outflows. Indeed, it has been shown that the 
asymptotic shape of steady axisymmetric magnetized outflows is  
either paraboloidal or cylindrical according to whether the 
electric current carried at infinity vanishes or not
(See Lery, Heyvaerts, Appl \& Norman 1999 and reference therein). 
In the present model, the axial electric current
carried by the flow is given by
\begin{equation}
I\propto\left({r}/{r_o}\right)^{\alpha+1/4},
\end{equation}  
with $1/4\ge \alpha> -1/2$.
If $\alpha<-1/4$ (\ie solutions with 
a gain in energy by radiation), the current is only large 
close to the protostar and will vanish at infinity. 
Thus the outflow should have a paraboloidal or conical shape.
On the other hand, for the special case $\alpha=-1/4$, 
\ie solutions with radiative balance, the current is not a
function of the distance to the central object but only a 
function of the angle. The current will have a finite value 
at infinity and therefore the flow should focus cylindrically. 
A similar conclusion is obtained for $\alpha>-1/4$ where the 
current increases with the distance. Consequently, for this 
range, the outflow should collimate cylindrically 
far from the source, at least in the axial regions. 

\begin{figure}
\psfig{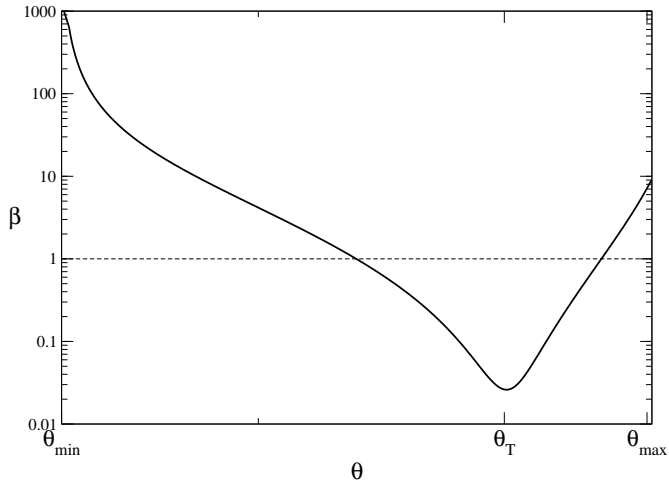}
\caption[ ]{
{\bf Variations of the plasma $\beta$ parameter
with the angle} along an inner streamline
starting from the cloud boundary for the circulation model. 
$\theta_{min}$, $\theta_{max}$ and $\theta_{T}$
respectively correspond to the angles at the minimum, the 
maximum and the turning point of the solution
($\alpha=-0.12$,$\Theta_0=0.5$).
\label{fig:beta}
}
\end{figure}
It is also interesting to study the plasma $\beta$ parameter
that corresponds to the gas to magnetic pressure ratio. For our model, 
$\beta$ is given  by $\beta={y_\phi^2 P}/{u_\phi^2}$.
This quantity varies drastically with the angle in the circulation model
as shown in Fig.~\ref{fig:beta}. $\beta$ starts from 10 in the
infall region, reduces to $10^{-2}$ at the turning point and finally
increases up to several hundreds in the jet region.

\subsection{Down to the Thin Disk}

\begin{figure}
\psfig{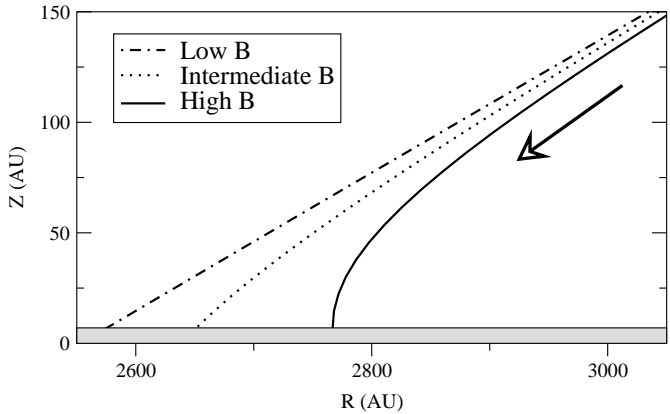}
\caption[ ]{
{\bf Shapes of poloidal sections of the 
magnetic/flow surfaces above the disk}
for solutions of the infalling envelope model
with different initial conditions for the 
the self-similar magnetic variable. 
The grey zone corresponds to the thin disk. 
In this solutions, the gas is flowing
down as indicated by the arrow.
(Same set of parameters as in Fig.~\ref{fig:Be5})
\label{fig:anchor}
}
\end{figure}
The last point that we shall discuss in this section concerns 
the boundary conditions at the interface with the thin disk,
that in the model, is regarded as an open boundary. Physically, 
this corresponds to the entry of the gas into this region,
 while it is still falling onto the central object. The gas 
is not effectively falling onto the thin disk since
self-gravity is not taken into account in the model.
We have represented the shape of the poloidal component of 
the magnetic field above the disk in Fig.~\ref{fig:anchor}
for several solutions, that differ only by 
 the initial conditions for the 
the self-similar magnetic variable. 
We find that the angle of the trajectory of the 
infalling gas when it goes out of the boundary directly
depends on the initial magnetic field. For example, 
nearly perpendicular field lines can be obtained for 
a large magnitude of the field. 
When the magnetic field decreases, the lines bend and 
present a larger angle w.r.t. the perpendicular to the 
disk as shown in Fig.~\ref{fig:anchor}. In this case, the gas
entering the disk has a non-vanishing radial velocity, and
the magnetic field has an helical structure.

\section{Discussion} \label{sec:discussion}

\subsection{The Molecular Outflow Acceleration}

From an observational point of view, molecular outflows
generally consist of a pair of oppositely-directed
and poorly collimated lobes, which are symmetrically located
about an embedded young stellar object, and detected in broad 
mm-wave emission lines, especially of carbon monoxide (CO) 
(\cite{Bachiller99}). They are usually moderately collimated 
(with initial opening angles $\theta_i\approx20^{\circ}-90^{\circ}$), but
highly collimated molecular outflows ($\theta_i<10^{\circ}$)
are also observed (\eg Andr\'e \etal 1990). 
The most luminous YSOs, which generally correspond to the most 
massive ones, have outflows with relatively wide-opening angles
(\cite{Churchwell}). Their outflow masses are estimated to be  
up to several hundred solar masses (\eg $200 M_\odot$ for 
Mon R2 (\cite{Tafalla})), and in general they 
can be more than 5 times larger 
than the final mass of the central star.

In the present model, most of the properties of molecular outflows
can be understood in terms of the jet and the 
outflow having separate origins, although they 
may interact at their interfaces.
Observationally, the spatial coincidence of shocked molecular 
hydrogen bows with peaks in the CO outflow emission suggest 
that the jet and outflow interact (\eg Davis \& Eisl\"offel 1995).
In our model, the jet comes from 
the inner-most part of the central object, \ie the inner 
accretion disk and the forming star, 
while the molecular outflow is produced by the circulation
and is tied to the physics of the molecular cloud. 
We self-consistently model the infall, outflow, and jet regions using the 
same set of equations around the same central object.  Even though the three 
different parts of the model are formally disconnected, 
they are related through strict boundary conditions at 
their interface surfaces.

This has interesting 
consequences concerning the subsequent interaction of 
the flow with the coaxial jet. Since in our case, the 
difference in velocities between the jet
and the molecular outflow material are reduced
from the start, the shocks in the zone of acceleration
due to their interaction should be less strong. 
The central fast jet has still the largest part 
of the total momentum per unit area, and the molecular outflow 
could undergo a prompt entrainment from the head of the jet. 
The effects of this entrainment, however, are reduced because 
of the motion induced in the CO outflow by the circulation 
around the protostar. 

But the most interesting feature of the circulation model
is probably that it could produce solutions where the mass of
the molecular outflow is larger than the final mass of 
the forming star. This would particularly true if self-gravity 
were included in the model. One might then understand how
bipolar outflows from massive protostars are observed to transport 
masses largely exceeding those of the associated stars.
For high luminosity objects, solutions can be obtained that show
well-collimated jets, even though the CO flow may
appear poorly collimated (\cite{Bachiller99}).

Other important features of the model are that the 
distribution of the flow mass with velocity  approximately follows 
a power law, and the flow collimation increases with 
velocity. The model also shows solutions in which the flow velocity 
increases with distance along a line parallel to the outflow axis.
Also, the synthetic CO spectra 
corresponding to these solutions nicely reproduce most of 
the observational features of molecular outflows (\cite{FH,LHF}). 
The acceleration is relatively gentle in this type of model,
so that molecules are advected with the flow with relatively 
little destruction. This is in contrast to the more severe 
acceleration that occurs when gas is entrained by a fast jet, 
which tends to destroy the molecules (\cite{Tafalla,downes}). 
In our model, several tens of solar masses are moving 
at supersonic velocities while remaining neutral and in 
molecular form. The calculations produce solutions where 
the outflow can have large  opening angles, and where the 
most massive protostars produce the 
fastest and the less collimated outflows (\cite{LHF}).

In conclusion, we suggest that CO outflows are dominated by the global 
circulation of material around the protostar, except for in
a thin layer surrounding the jet, where the dynamics is governed
by entrainment. We stress that, in the present model, the two flows 
(atomic and molecular) are not strongly linked dynamically, and,
hence, there is no need to transfer large momentum from the jet
to the molecular outflow through the entrainment processes.

\subsection{Molecular Cavities}

At CO mm-wavelengths, \eg L1551 (\cite{Moriarty,Hartigan}), 
NGC 2071 (\cite{Moriarty2}), or Mon R2 (\cite{Tafalla}) 
and also in the infrared band (\cite{Close478}), 
cavities inside molecular outflows are observed. 
The edges of such cavities, located between the jet
and the molecular outflow, are delineated by $H_2$ emission.
Indeed, close to the source, ``throats'' are observed 100 to 200 AU 
above the accretion disk (\cite{Close489}).
The association of outflows with cavities is not limited
to low-luminosity sources (\cite{Campbell,Tafalla}). For example, 
the blue lobe of AFGL 490 (\cite{Campbell}) coincides with a conical 
nebula of scattered light, suggesting that the outflow 
interacts with a shell of dense gas surrounding a partly 
evacuated channel.

The cavities can have large opening angle 
in the range 90$^{\circ}$-125$^{\circ}$ 
(\eg Velusamy \& Langer 1998), and their nature is unclear
(\cite{Chandler96,Gomez,Gueth}). 
These observations are in good agreement with the opening angles in our 
model and in wide-angle wind models, which produce wider 
cavities than jet models.
It has also been proposed that
the cavities could be produced by the wandering of a precessing jet. 
However, it appears that the dynamical 
time-scale of the jet is too short  for this explanation to 
work in some observed sources (\cite{Chandler96}).

In the present model, the molecular cavities   
may be identified with the substantial decrease in density  
in the intermediate region between the jet and the 
molecular outflow (See Fig.\ref{fig:density} in the axial region).  
Note that solutions with wider opening angles than the one 
presented here can easily be obtained.
Moreover, the central jet region may be comprised mainly of atomic gas, which 
is not observable in molecular lines.  This is likely because the gas
will probably be dissociated and ionized near the protostar.  
As the gas escapes
and cools, atomic gas will form rapidly.  However, 
the timescale for the production
of molecules is much longer. Thus, the molecular outflows 
appear as a hollow conical structure (\cite{LHF}).
The model indicates a wide range of opening angles 
from 40$^{\circ}$ to 140$^{\circ}$, depending on the 
input parameters, in agreement with the observed values. 
However,
this angle may vary with time but the timescale for the widening 
is expected to be much longer than the typical dynamical timescale. 
Thus our model may represent a quasi-steady flow pattern at some 
instant in a slowly changing outflow. Anyhow, the cavity is a 
result of the circulation itself and we do not need to assume that 
the jet is precessing.

\subsection{Asymmetric Outflows}

\begin{figure}
\psfig{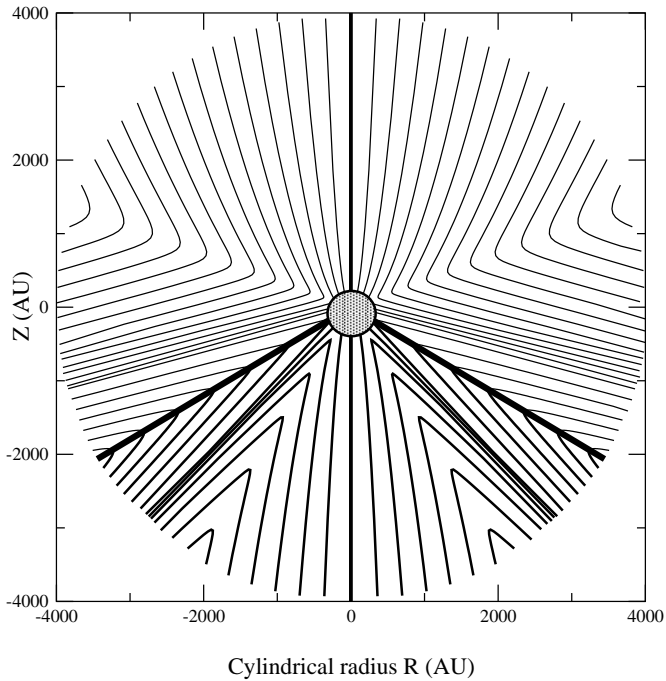}
\caption[ ]{
{\bf Streamlines of solutions corresponding to asymmetric outflows}
with quadrupolar geometry. The parameters are similar to
the previous solutions ($\alpha=-0.12$,$\Theta_0=0.5$)
but the maximum opening angles of 
the solutions are not limited to 90$^{\circ}$. 
\label{fig:asym}
}
\end{figure}

Another important observational issue is the large number 
of bipolar optical outflows showing properties
that differ substantially in the blue-shifted and 
red-shifted parts (\cite{hirth,Bacciotti99}). 
In the axisymmetric quadrupolar circulation model, solutions  
are not limited to an hemisphere, and consequently  
can differ from side to side of the disk (\cite{FH,LHF}). . 
We present an example of such a solution in Fig.~\ref{fig:asym}.
In this figure, the streamlines are represented for all the three 
different parts of the model, \ie the circulation model, the 
jet model and the infall envelope. The solutions  have the same 
set of parameters on both sides of the equator, \ie 
$\alpha=-0.12$,$\Theta_0=0.35$, $r_o \approx 100$AU, $M\approx 1M_\odot$,
but have different opening angles, that we prescribe as 
input parameters ($\theta_{max}=2\pi/3>\pi/2$ for the upper part, 
and $\theta_{max}=\pi/3<\pi/2$ for the lower part).  
These solutions may be generated by different
conditions on both sides of the disk, for example in a 
region of a cloud that had a strong pre-existing density gradient,
or at the edge of a molecular cloud.
The latter case is similar to the standard interpretation 
of the asymmetric objects as an outflow breaking out of 
the cloud on one side. 
Note, however, that contrary to the optical jet case, for 
the molecular outflow the weak lobe will be pointing out
of the cloud as observed for many unipolar CO outflows known, such as 
NGC2024-FIR5 (\cite{Richer92}), or the almost-unipolar 
HH46-47 system (\cite{Chernin91}).
Our model also predicts that the 
dominant outflow lobe will be on the side pointing into 
the cloud because that is the side where material can be gathered.
Thus, the various asymmetries observed on opposite sides 
of many bipolar outflows,
could find a natural explanation in the quadrupolar geometry. 
In any event, detailed comparisons of solutions for a specific 
asymmetric source could be a good test of our hypothesis.
It is also noteworthy that purely unipolar 
models had already been found in the context of the circulation model 
when treated without Poynting flux (\cite{Fiege}).

\subsection{Collimation}\label{sec:coll}

Several mechanisms, from purely hydrodynamic (\cite{MF}) to 
magnetic ones, have been proposed to account for the collimation 
of jets. Magnetic processes seem very promising and they are 
generally accepted as the primary option. In this case, the 
outflow is collimated by the pinch effect of the toroidal 
component of the magnetic field that arises as a consequence 
of the rotation of the disk (\cite{lery2,Breitmoser}). 
In cylindrical jet asymptotics, the outflowing plasma reaches
an equilibrium wherein it is confined by magnetic forces or 
gas pressure gradients, while it is supported by centrifugal 
forces or gas pressure gradients. 
The idea that the external medium has, in fact, a non-negligible 
effect has been raised recently (\cite{Okamoto,leryfrank}).
Indeed, magnetized winds may not cylindrically 
collimate without an external help, such as the
channeling effects of thick accretion discs and/or 
confinement resulting from the ambient medium. 
Moreover, in spite of the success of magnetic models, 
recent numerical studies have shown that pure 
hydrodynamic collimation could be effective at 
producing jets (\cite{Icke,MF}), thanks to the ``shock 
focusing'' mechanism due to the interaction of the 
outflow with the external environment. 
In the present model, the collimation of the axial jet arises to a 
large extent from the hoop stress of the toroidal field 
component but it is also reinforced by the density gradients 
in the surrounding molecular outflow.
Thus, both hydrodynamic and magnetic effects play a role. 
Note that the present model predicts that the collimation
does tend to increase with the velocity of the gas
regardless of the mass of the central object (\cite{FH,LHF}).

The presence of low density polar regions may also
give a preferential direction and help the alignment 
of jets coming from a binary system, as in the case of L1551.
Here,  HST imaging and spectroscopy from the ground (\cite{Fridlund,Bo})
indicates the presence of two individual jets almost aligned 
(angle between the jets less than 20$^{\circ}$) and lying inside a CO cavity.

\subsection{Possible origin of the quadrupolar topology}

Most of current outflow models are based on a dipolar geometry
where the magnetic field lines are either advected via accretion 
towards the central object in the accretion disk or are created in situ.  
Aburihan et al. (2001) proposed a simple, qualitative mechanism to 
generate a quadrupolar field topology, which we briefly summarize here.  
The basic idea is that advection deforms the field into a predominantly 
quadrupolar topology as the magnetized gas collapses.
During the earliest stages of formation, the magnetic field near 
the protostar would be mainly dipolar as the gas starts to collapse.  
Since the magnetic field is frozen into the gas,
the infalling material advects the magnetic field toward the 
accreting core until it is strongly pinched toward the protostar
in the equatorial region.  At this point, the field is mainly radial.
Poloidal pressure gradients, with both radial and angular components, 
develop as the protostar forms.  Eventually these pressure gradients 
become strong enough to deflect some of the gas toward the polar 
regions (\cite{LHF}).  The magnetic field near the protostar
is advected with this outflowing gas, and eventually forms magnetic 
arches near the polar axis.  A predominantly quadrupolar field 
topology remains as these magnetic arches are advected to large radii.  
Other mechanisms might exist to convert an initially 
dipolar field into a quadrupolar 
geometry, but these might rely on reconnection or dynamos.

\section{Conclusion} \label{sec:concl}

We have proposed a new global MHD model for flows around YSOs.
The global model combines a jet model, a circulation model and 
an infalling envelope model. Instead of the usual mechanisms 
invoked for the origin of molecular outflows, the 
outflow is powered by the infalling matter through a heated 
quadrupolar circulation pattern around the central object. 
The molecular outflow may still be affected by 
entrainment from the fast jet, but this would be limited to the
polar regions and it would not be the dominant factor behind 
its acceleration.

The solutions are developed within the context of 
$r$-self-similarity. We use the set of steady 
axisymmetric MHD equations for all the three parts of the model. 
The only physical scales that enter 
into our calculation are the gravitational constant $G$, the 
fixed central mass $M$, and a fiducial radius $r_o$.
The parameters of the model are 
the indexes $\alpha$ and $\alpha_f$ of the self-similar system,
and $\theta_{min}$ and $\theta_{max}$, the limiting angles 
between the three regions of the model.
Magnetic field and streamlines are required to be quadrupolar 
in the poloidal plane for the circulation model.

We have explored the physical behaviour of the model
by studying the flow through its various energy components along 
streamlines.
We have reported here solutions for the 
three different parts of this self-similar model. An example
of solution is presented for a one solar 
mass object, even though this class of model can apply to both low 
and high mass protostars. 
The solutions show dynamically significant density 
gradients in the axial region, precisely where the radial 
velocity and collimation are the largest. 
Rotation is maximum at the turning point but it is still 
dynamically  important close to the disk.

Our model can reproduce mass loss rates that are in agreement with values 
derived from observations. For the specific case that is 
presented here, we obtain 
$\dot{M}_{jet} /\dot{M}_{acc} \sim 0.1 $, as observed
in several systems, even if our model does not deal with 
the precise jet launching mechanism.  Other values, typically ranging from 
$\sim 0.04$ to $\sim 0.4$ are possible, which 
may impose additional observational 
constraints on the parameter space of our model.

From an observational point of view,
we clarify the nature of the molecular outflow acceleration 
and its relation with the fast jet, by providing a global 
picture of the jet/outflow system, which does not primarily 
rely on entrainment (prompt or turbulent). 
In this scenario, the observed molecular cavities 
result from the circulation pattern itself, and there is no need 
for jet wandering. Our model helps to explain 
the presence of high mass molecular outflows around massive 
objects, being the molecular material simply circulating around 
the central object from the nebula to the bipolar outflows. 
The various 
asymmetries observed on opposite sides of many bipolar outflows
could find a natural explanation in the quadrupolar geometry;
our model predicts that the dominant outflow lobe will 
be on the side pointing into the cloud.
The present study suggest
that radiative heating and the Poynting flux may ultimately be
the main energy sources driving molecular outflow.
Finally, although the details of the jet mechanism may be peculiar
to individual objects we believe the infall-outflow circulation
to arise naturally given accretion, and thus could also
be present in AGN's and around suitably placed objects
such as SS433.    

\begin{acknowledgements}
TL and AF acknowledge support from NSF Grant AST-0978765 and by the University 
of Rochester's Laboratory for Laser Energetics.  JDF acknowledges support from 
CITA and NSERC fellowships. 
\end{acknowledgements}


\end{document}